# On elementary particles spectra within the framework of curvilinear waves electrodynamics

Alexander G. Kyriakos

*Saint-Petersburg State Institute of Technology,
St. Petersburg, Russia.*

*Present address: Athens, Greece, e-mail:* agkyriak@yahoo.com

**ABSTRACT**. In previous papers within the framework of the non-linear electromagnetic field theory - curvilinear waves electrodynamics (CWED) - we have considered the opportunity of occurrence of electromagnetic elementary particles, similar to leptons and hadrons. In present paper we will show that on this basis the spectra of more complex electromagnetic particles can also be naturally created

## 1.0. Introduction

According to modern representations, all elementary particles are the bound states (including the exited states) of a small set of particles. For example, according to (Gottfried and Weisskopf, 1984): "The nucleon is simply a basic state of a compound spectrum of particles which we have named a baryon spectrum. Similarly pion is the lowest state of meson spectrum".

In present paper we will show that in the framework of CWED the spectra of more complex electromagnetic particles as bound and exited states of a small set of some basic particles, can be formed.

### 1.1. Electromagnetic elementary particles of CWED

As elementary particles of CWED we mean the electromagnetic particles that arise and are described within the framework of CWED. As we have shown, it is possible to compare them with the actually existing elementary particles described by the quantum field theory, whose existence is confirmed with experiments (Kyriakos, 2004a; b; c).

Generally each elementary particle is defined by a set of various characteristics: a mass, a spin, an electric charge, the strong and weak "charges" (i.e. the characteristics, which define intensity of strong and weak interaction), the numbers of "affinity" (numbers, owing to which one family of particles differs from another - lepton, baryon and other numbers), etc.

The particles characterized by identical characteristics, except for any one of them, create a spectrum of elementary particles regarding this variable characteristic. For example, if as such variable characteristic the mass of particles is accepted, they speak about a mass spectrum of elementary particles.

According to the modern theory there are some limiting conditions of the composition of elementary particles, which can be named the conservation laws of this characteristic: e.g. the laws of conservation of energy, momentum, angular momentum, laws of conservation of an electric charge and charges of other interactions, laws of conservation of numbers of "affinity", etc. Some laws (principles) also exist, such as a principle of uncertainty of Heisenberg, which restrict the transition from one family or a spectrum of particles to another.

As is known, the existing field theory cannot explain the occurrence of elementary particle characteristics and cannot deduce the majority of conservation laws of these characteristics: they are entered as consequences of experiments.

If to speak, for example, about mass spectra of particles, the following restrictions exist on creation of such spectra:

1) according to the energy-momentum conservation law the rest free light particles cannot break up to heavier particles, but heavy particles can break up to more light particles;

2) nevertheless, according to a uncertainty principle of Heisenberg, heavy particles cannot comprise the light particles as a ready particles (for example, the neutron cannot comprise electron as a free particle).

The conclusions of the quantum theory are undoubtedly correct and was confirmed by experiments, and we should show, that they do not contradict to the results of CWED.

## 2.0. A hypothesis of formation of spectra of elementary particles in CWED

Within the framework of CWED the electromagnetic twirled waves (EM-particles) possess the same characteristics, as quantum elementary particles. As we saw (Kyriakos, 2004a,b,c), the twirled harmonic waves, appearing here, can have integer or half spin, can be charged or neutral, etc. The mass of particles within the frameworks of CWED is the "stopped" energy of the twirled standing wave. Thus, roughly speaking, to a heavy particle by our representation corresponds the twirled wave of high frequency, and to light particle - the twirled wave of lower frequency. Thus, we should explain the existence of spectra of the particles relatively to all these particularities.

To the simple harmonic waves in Classical Electrodynamics (briefly CED), the twirled harmonic waves in CWED correspond. Does exist in CED the opportunity of coexistence of several waves as some material formation - an elementary particle, in which the characteristics of various waves can be superposed?

As we know, such opportunity actually exists and it consists in the waves superposition, which leads to various forms of coexistence of normal harmonic waves and to the occurrence of complex non-harmonic waves, which "consist" of harmonic waves of various frequencies.

Analogically to the representations of classical theory of EM waves, whose non-linear generalization our theory is, we assume that:

*the reason of complication of EM particles and of occurrence of its spectra is the superposition of simple (harmonic) twirled waves, and the reason of disintegrations of particles is the disintegration of the compound twirled waves.*

The purpose of our paper will be to show that such superposition exists and its description completely corresponds to modern theoretical representations and is in full accordance with the experimental data.

As CWED is non-linear generalization of classical (linear) electrodynamics, it is possible to assume, that the opportunity of the mathematical description of creation of waves spectra should exist already in CED. Besides, since mathematical description of CWED completely coincides with the mathematical description of quantum electrodynamics (QED), we should show that the similar forms exist in QED as well as in CWED. Therefore we will build the further statement on the comparison of these three areas of the description of elementary wave constituents of the nature: CED, CWED and QED.

### 2.1. Superposition of «linear» waves

As is known (Grawford, 1970), the motion of compound system can be represented by superposition of more simple motions, occurring simultaneously and named "modes" (terms: "simple harmonic oscillation", "harmonics", "normal oscillation", "own oscillation", "normal mode" or simply "mode" are identical). Properties of each mode of any compound system are very similar to properties of simple harmonic oscillator.

In many physical phenomena the system motion represents a superposition of two harmonic oscillations, having various angular frequencies $\omega_1$ and $\omega_2$. These oscillations can, for example, correspond to two normal modes of the system, having two degrees of freedom. (a known example of such system is the molecule of ammonia (Grawford, 1970)).

It is true as well for the quantum mechanical waves, described by quantum wave functions. It is possible to illustrate this fact on an example of formation of an energy spectrum of electron in an orbit of hydrogen atom. Really, the electron energy spectrum in an electron-proton system is from the general point of view a spectrum of electron masses. It is possible to speak about a basic mass (basic energy) in not exited state, and about a lot of masses of electron in the exited states, when electron receives additional portions of energy (mass). These portions are very small in comparison with the basic electron energy (mass), which we cannot say about masses of the particles, appearing at disintegrations of elementary particles. But, nevertheless, it does not exclude that these are the phenomenon of the same type. The increase of electron mass occurs due to absorption of photons, and the reduction of mass takes place due to emission of photons. On the other hand, we actually cannot tell, that the electron contains a photon as a ready particle.

It is also easy to show (Grawford, 1970), that the change of electron energy as a result of its excitation by a photon corresponds to a hypothesis about the occurrence of new particles owing to superposition of waves.

Let's consider the steady-states of the electron in one-dimensional potential well with infinitely high walls, whose coordinates are $z = -\frac{L}{2}$ and $z = +\frac{L}{2}$. We will also assume that the electron state is defined by superposition of the basic state and the first exited state:

$$\psi(z,t) = \psi_1(z,t) + \psi_2(z,t), \tag{2.1}$$

where $\psi_1(z,t) = A_1 e^{-i\omega_1 t} \cos k_1 z$, $k_1 L = \pi$, $\psi_2(z,t) = A_2 e^{-i\omega_2 t} \sin k_2 z$, $k_2 L = 2\pi$.

The probability of electron existence in the position $z$ in the time moment $t$ is equal to

$$\begin{aligned}|\psi(z,t)|^2 &= |A_1 e^{-i\omega_1 t} \cos k_1 z + A_2 e^{-i\omega_2 t} \sin k_2 z|^2 = \\ &= A_1^2 \cos^2 k_1 z + A_2^2 \sin^2 k_2 z + 2 A_1 A_2 \cos k_1 z \cdot \sin k_2 z \cdot \cos(\omega_2 - \omega_1)t\end{aligned}, \tag{2.2}$$

We can see that the probability expression has a term, which makes harmonic oscillations with beats frequency between two Bohr frequencies $\omega_1$ and $\omega_2$. The average electron position in space between the wells can be found by means of the expression:

$$\bar{z} = \frac{\int z |\psi|^2 dz}{\int |\psi|^2 dz} = \frac{32L}{9\pi^2} \frac{A_1 A_2}{A_1^2 + A_2^2} \cos(\omega_2 - \omega_1)t, \tag{2.3}$$

where the integration is from one wall $-\frac{L}{2}$ up to the other $+\frac{L}{2}$.

Obviously, the frequency of radiation is defined by beats frequency. Actually, electron is charged and, consequently, it will emit out of the electromagnetic radiation of the same frequency, with which it oscillates. From the equation (1) we see, that average position of a charge oscillates with beats frequency $\omega_2 - \omega_1$. Therefore the frequency of radiation is equal to beats frequency between two stationary states:

$$\omega_{rad} = \omega_2 - \omega_1, \tag{1.4}$$

In the framework of CWED, the non-normalized quantum wave function is simply the wave field. As a consequence of this fact, the square of this wave function (i.e. the possibility density in the framework of QED) is the energy density. As example of such problem in framework of CWED we will consider the calculation of more general case of the interference between waves of various frequencies.

We will assume, that we have two EM waves 1 and 2, having electric fields $\vec{E}_1$ and $\vec{E}_2$. The full field in the fixed point P of space will be the superposition of $\vec{E}_1$ and $\vec{E}_2$. Using complex representation of oscillations, we will write the expression for superposition of oscillations:

$$\vec{E}(t) = E_1 e^{-i(\omega_1 t + \varphi_1)} + E_2 e^{-i(\omega_2 t + \varphi_2)}, \qquad (2.5)$$

The energy flux is proportional to average value of $\vec{E}^2(t)$ for period T of the "fast" oscillations, occurring with average frequency:

$$\begin{aligned} 2 <E^2(T)> &= |E(t)|^2 = \left| E_1 e^{-i(\omega_1 t + \varphi_1)} + E_2 e^{-i(\omega_2 t + \varphi_2)} \right|^2 = \\ &= E_1^2 + E_2^2 + 2 E_1 E_2 \cdot \cos[(\omega_2 - \omega_1)t + (\varphi_1 - \phi_2)] \end{aligned} \qquad (2.6)$$

As we see, the energy flux varies with relatively slow beats frequency $\omega_2 - \omega_1$.

## 3.0. Superposition of the twirled electromagnetic waves

So, we should show at first that the superposition of the twirled electromagnetic waves exists and secondly that owing to it, it is possible to receive all those results, which are known from the theory of linear electromagnetic waves. In other words, it is necessary to show, that in this case there are actually series (spectra) of particles, each of which represents complication due to the superposition of other twirled waves.

As is known, all the phenomena of superposition of waves and their disintegration are described by Fourier theory (Fourier analysis-synthesis theory), in which it is shown, that any field can be synthesized from harmonic waves or analysed to harmonic waves. We will show that Fourier theory is true in case of the twirled waves as well as in case of linear waves.

### 3.1. The real and complex form solutions of the wave equation, as reflection of an objective reality

As is known, the wave equation

| (CED form) | (CWED form) |
|---|---|
| $\left( \dfrac{\partial^2}{\partial t^2} - c^2 \vec{\nabla}^2 \right) \vec{\Phi}(y) = 0$, where $\vec{\Phi}(y) = \{E_x, E_z, H_x, H_z\}$ | $\left[ (\hat{\alpha}_o \hat{\varepsilon})^2 - c^2 (\hat{\vec{\alpha}} \, \hat{\vec{p}})^2 \right] \Phi = 0$, where $\Phi = \begin{pmatrix} E_x \\ E_z \\ iH_x \\ iH_z \end{pmatrix}$, $\hat{\varepsilon} = i\hbar \dfrac{\partial}{\partial t}$, $\hat{\vec{p}} = -i\hbar \vec{\nabla}$ and $\hat{\alpha}_0$; $\hat{\vec{\alpha}}$; $\hat{\beta} \equiv \hat{\alpha}_4$ are Dirac's matrices |

has the solution, which can be written down in the form of real periodic (in particular, trigonometric) functions, as well as in the form of complex (in particular, exponential) functions

| $\vec{\Phi}(\vec{r},t) = \vec{\Phi}_0 \cos(\omega t - \vec{k} \cdot \vec{r})$ | $\Phi = \Phi_o e^{-i(\omega t \pm ky)}$ | or |
| $\vec{\Phi}(\vec{r},t) = \vec{\Phi}'_0 \sin(\omega t - \vec{k} \cdot \vec{r})$ | $\begin{cases} \vec{E} = \vec{E}_o e^{-i(\omega t \pm ky)}, \\ \vec{H} = \vec{H}_o e^{-i(\omega t \pm ky)}, \end{cases}$ | |

Nowadays it is considered that the representation of the wave equation (or oscillation equation) solution in complex form is only a formal mathematical method, since the final solutions should be real. It was also marked, that the use of complex representation is dictated only by the reasons of convenience, as long as in many cases the mathematical operations with exponential functions are carried out much easier, than with trigonometric.

We have shown (Kyriakos, 2005), that within the framework of CWED the exponential solutions have an actual meaning, if we understand them in geometrical sense as the description of motion of a wave on a curvilinear (particularly circular) trajectory (or, in other words, as the sum of two linear mutual-perpendicular oscillations). (We have noted that due to this fact the solutions of the wave equations of the quantum theory are not real, but complex wave functions).

*Thus, it is possible to assume, that the existence of the real and complex solutions of the wave equation indicates the existence in the nature of two types of real objects: the linear and twirled (curvilinear) waves. At that the real functions describe linear waves, and the complex functions describe the curvilinear (twirled) waves*

As is known, the functions, describing complex periodic and non-periodic processes of non-harmonic type can be presented by the sum of harmonic functions owing to Fourier analysis-synthesis theory. It must be noted that the Fourier analysis-synthesis theory allows working equally both with real and complex functions.

*From this the extremely important conclusion follows that all tools of the Fourier analysis-synthesis theory of functions in complex representation is the mathematical apparatus, describing the superposition and decomposition of complex twirled waves (i.e. of complex electromagnetic elementary particles).*

In other words, the complex representation of electromagnetic waves and all mathematical apparatus of Fourier analysis-synthesis theory represent mathematical tool of CWED in the same degree as the mathematical apparatus of the real functions of Fourier analysis-synthesis theory represents the mathematical tool of usual linear Maxwella-Lorentz theory (i.e. CED).

*Due to above, the non-linear theory of the twirled waves is the theory in which the principle of superposition takes place as well in the linear theory.*

For this reason the Maxwell-Lorenz theory can be written down in a complex form and it looks in such form simple and consistent. Transition from the twirled waves to linear (i.e. to one of components of the twirled wave) corresponds to transition from complex values to real (which is carried out by calculation of a real part of complex value of quantities and functions).

Let us consider now some details of the Fourier analysis-synthesis theory in case of superposition of the twirled waves.

## 4.0. Elementary particles as wave packets

As is known, in case of superposition more than two linear harmonic running waves the wave groups or wave packets are formed, which are the limited in space formations, which transfer energy and move with some group speed.

In the quantum mechanics a wave packet (Physics encyclopedia, V.1, 1960) is the concept, designating a field of waves of a matter, concentrated in the limited area. The probability to find a particle is distinct from zero only in the area, occupied by a wave packet. The less are the sizes of a packet, the more is the particle localized. It is possible to consider this wave field as result of superposition of the certain set of plane waves (from here follows the name "a wave packet"), corresponding to the different wavelengths and, hence, to the different momentums of a particle. The possibility of similar decomposition on flat waves is a simple result of a possibility to analyse any function in a Fourier series or Fourier integral.

It is only meaningful to apply the concept of a wave packet when the wave numbers $\vec{k}$, used in it, are grouped near to some $\vec{k}_0$ with small variation $\Delta\vec{k}$, $\Delta k \ll k_0$. In this case the wave field (i.e. wave packet) during significant time will move as a whole, being a little deformed, with the group speed $k_0 u = \left(\dfrac{d\omega}{dk}\right)_{k=k_0}$ corresponding to a speed of a particle, described by this wave packet. As is known, the smearing of the wave packet does not take place if it can be decomposed on standing waves, i.e. if in the decomposition series for each vector $\vec{k}$ the vector $-\vec{k}$ with the same amplitude is also entered.

Thus, knowing that superposition of linear waves leads to formation of the linear wave packets, it is consistent to conclude that superposition of the twirled waves leads to formation of the twirled wave packets, i.e. to the compound electromagnetic elementary particles.

It is characteristic that the representation of wave function by the Fourier sum, i.e. Fourier series (in case of periodic function) or by the Fourier integral (in case of non-periodic function):

| Real form: $f(t) = \dfrac{a_0}{2} + \sum_{n=1}^{\infty}(a_n \cos n\omega\, t + b_n \sin(\omega\, t))$, where $a_n, b_n$ are the Fourier coefficients. | Complex form: $f(t) = \sum_{n=-\infty}^{\infty} c_n e^{-in\omega\, t}$, where $c_n$ are the Fourier coefficients. |
|---|---|

contains the negative frequencies, which in the linear theory have no a place. As is known (Matveev, 1985), in classical optics they take into consideration that $e^{i\omega t}$ describes the complex unit vector, which is started from the origin of coordinates, and which at increase of time *t* rotates around this origin in a positive direction (by a rule of the right screw). In the same time the complex unit vector $e^{-i\omega t}$ rotates in the negative direction.

Thus, the reference to the negative frequencies is connected with change of basic functions, by means of which Fourier-transformation is carried out, namely with transition to the rotating complex vectors as to the basic functions of Fourier-transformation. The above completely corresponds to our representations on conformity of Fourier mathematical tools with the requirements of CWED.

As a simple example of formation of a wave packet, we will consider a packet formed by the equidistant rectangular frequency spectrum of waves of equal amplitudes. The description of superposition of such waves can be made in real (Grawford, 1970) as well as in a complex form (Matveev, 1985), that reflects the existence of the linear and non-linear world of particles.

We will find the exact expression for an packet $\psi(t)$ formed by superposition of *N* various harmonic components, which have equal amplitude *A*, an identical initial phase (equal to zero) and the frequencies distributed by regular intervals between the lowest frequency $\omega_1$ and the highest frequency $\omega_2$. Generally we have:

| (real form) $\psi(t) = A\cos\omega_1 t + A\sum_{n=1}^{N-1}\cos(\omega_1 + n\delta\omega)\, t + A\cos\omega_2 t$ | (complex form) $\psi(t) = A\sum_{n=0}^{N-1} e^{i(\omega_1 t + n\delta\omega t)}$ |
|---|---|

where $\delta\omega$ is the frequency, on which two next components differ, and $n = 1,2,3,...,N-1$ and $\omega_2 = \omega_1 + N\delta\omega$.

This formula represents the complex wave function in the form of linear superposition of the lot of strictly harmonic components. It appears, that this sum can be expressed in the form, which are the generalization of a case for two oscillations:

$$\psi(t) = A(t)\cos\omega_m t, \qquad (4.1)$$

where $A(t) = A\dfrac{\sin(0{,}5N\delta\omega \cdot t)}{\sin(0{,}5\delta\omega \cdot t)}$ is the variable amplitude, $\omega_m = \dfrac{1}{2}(\omega_1 + \omega_2)$ is the average frequency of a wave packet. The amplitude $A(t)$ describes a wave packet envelope. It is possible to show (Grawford, 1970), that for a wave packet, Heisenberg uncertainty principles are true, what proves their wave origin.

Since the twirled waves by their origin already represent the limited objects, it is possible to assume, that the electromagnetic particles should be combined not from infinite Fourier series, but they should be presented by the sum of the limited number of harmonics, i.e. of the twirled waves.

To describe the synthesis of the complex particles (packets) from more simple sub-packets, we will show, that any wave packet can be presented in the form of the sum of wave sub-packets. In this case, obviously, superposition (interaction) of several big packets can be considered not as superposition (interaction) of their separate harmonic components, but as superposition of their sub-packets.

Let's consider the splitting of a big packet into two sub-packets. We will present a compound wave $\psi(t)$ (see above (4.1)) in the following form:

$$\psi(t) = A\cos\omega_1 t + A\sum_{n=1}^{N-1}\cos(\omega_1 + n\delta\omega)t + A\cos\omega_2 t =$$

$$= (A\cos\omega_1 t + A\sum_{m=1}^{N_1-1}\cos(\omega_1 + m\delta\omega)t + A\cos\omega_2' t) + \quad , \quad (4.2)$$

$$+ (A\cos\omega_1' t + A\sum_{l=1}^{N_2-1}\cos(\omega_1' + l\delta\omega)t + A\cos\omega_2 t)$$

where $N = N_1 + N_2$, $\omega_2' = \omega_1 + N_1\delta\omega$, $\omega_1' = \omega_1 + (N_1+1)\delta\omega = \omega_2' + \delta\omega$.

Thus, we can represent the wave packet $\psi(t)$ as two sub-packets:

$$\psi(t) = \psi_1(t) + \psi_2(t), \quad (4.3)$$

where
$$\psi_1(t) = A\cos\omega_1 t + A\sum_{m=1}^{N_1-1}\cos(\omega_1 + m\delta\omega)t + A\cos\omega_2' t$$

$$\psi_2(t) = A\cos\omega_1' t + A\sum_{l=1}^{N_2-1}\cos(\omega_1' + l\delta\omega)\, t + A\cos\omega_2 t .$$

It is convenient to enter a contraction for a normal harmonic $\psi(t)$, and for a packet of waves $^\Sigma\psi(t)$, where sigma means the sum of harmonic waves (in particular, a sub-packet). Then representation of a packet in the form of the sum of sub-packets can be written down in the form of:

$$^\Sigma\psi(t) = {}^\Sigma\psi_1(t) + ... + {}^\Sigma\psi_2(t) = \sum_i {}^\Sigma\psi_i(t), \quad (4.4)$$

From the above-stated calculations it is visible that decomposition on sub-packets is not unambiguous, since the sub-packets can be grouped from harmonic waves in various ways. It is possible to assume, that the decay of the same particle on different channels can be considered as an opportunity of disintegration of a wave packet of the twirled waves on various sub-packets (i.e. the various sums of partial wave packets).

Using the above-stated reason it is easy to prove also that superposition (interaction) of sub-packets leads to the same consequences as interaction of separate harmonic waves, i.e. it leads to beats and to change of the energy level, independent from other non-interacting sub-packets (and also to the change of the energy level of the basic packets)

Except for curvilinearity in CWED there is one more serious difference from linear electrodynamics: in CWED alongside with the full periodic twirled waves (bosons), exist also the half-period twirled waves (fermions). This creates a great number of additional variants of the wave superposition, which are not present in linear electrodynamics. Besides, the curvilinearity enters into the physics one more characteristic of particles - the currents.

It is not difficult to understand that the superposition of the twirled waves in comparison with the superposition of linear waves has more variants in a spatial arrangement of waves, and, hence, more complex mathematical description. Actually we can see this in the case of description of hadrons (Kyriakos, 2004c).

It is easy to see, that the principle of superposition does not provide stability or, at least, metastability of compound electromagnetic particles. Thus, we should additionally find out the conditions of stability of the twirled waves.

## 5.0. The resonance theory of stability of elementary particles in CWED

As electromagnetic particles represent the spatial formations, here it is necessary to speak about spatial packets, which are formed by superposion of waves of a various direction in space (in the elementary case, by superposion of the harmonic waves propagating on three mutual-perpendicular coordinate axes).

As is known (Shpolskii, 1951), at the superposition of harmonic waves are formed the Lissajous figures of two various types. At commensurable frequencies of waves, are formed the standing waves; at incommensurable frequencies the motion of waves is refered to as quasi-periodic.

In the physics of waves and oscillations exist two sorts of the problems leading to the occurrence of the compound waves and oscillations.

An example of first type of problems is oscillation of the body volume (sphere, cylinder, torus, etc.), by which we represent a particle. Here the suitable mechanical example is the oscillation of the sphere prepared from a hydrophobic liquid, placed in water (for example, a sphere from mineral oil in water). In a microcosm the object, which possesses similar oscillations, is the drop model of a nucleus.

Problems concerning the same type are also the problems of oscillation of vortical rings in a perfect liquid or gas, studied by Kelvin (we will name conditionally such problems Kelvin's problems). In case of the oscillations of the linear vortex considered in work (Lord Kelvin, 1867) he obtains the exact solution. Here Kelvin has compared the radiation spectra of the atoms, obtained little time before by Bunsen, to possible spectra of oscillation of vortex. Comparison of such type of oscillations with observable results is available e.g. in works (Paper collection, 1975) and (Kopiev and Chernyshev, 2000). (It is necessary to note, that in his articles Kelvin used the term "atoms" in sense of Democritus as the smallest indivisible constituents, i.e. in modern terminology as elementary particles).

Certain of the Kelvin significant conclusions from the paper "Atom as Vortex" we cite below:

"The author called attention to a very important property of the vortex atom. The dynamical theory of this subject require that the ultimate constitution of simple bodies should have one or more fundamental periods of vibration, as has a stringed instrument of one or more strings.

As the experiments illustrate, *the vortex atom has perfectly definite fundamental modes of vibration*, depending solely on that motion the existence of which constitutes it. The discovery of these fundamental modes forms an intensely interesting problem of pure mathematics. Even for a simple Helmholtz ring, the analytical difficulties, which it presents, are of a very formidable character. The author had attempted to work it for an infinitely long, straight, cylindrical vortex. For this case he was working out solutions corresponding to every possible description of infinitesimal vibration.

One very simple result, which he could now state is the following. Let such a vortex be given with its section differing from exact circular figure by an infinitesimal harmonic deviation of order $i$. This form will travel as waves round the axis of the cylinder in the same direction as the vortex rotation, with an angular velocity equal to $(i-1)/i$ of the angular velocity of this rotation. Hence, as the number of crests in a whole circumference is equal to $i$, for an harmonic deviation of order *i* there are *i*-1 periods of vibration in the period of revolution of the vortex. For the case *i*=1 there is no vibration, and the solution expresses merely an infinitesimally displaced vortex with its circular form unchanged. The case *i*=2 corresponds to elliptic deformation of the circular section; and for it the period of vibration is, therefore, simply the period of revolution. These results are, of course, applicable to the Helmholtz ring when the diameter of the approximately circular section is small in comparison with the diameter of the ring, as it is in the smoke-rings exhibited to the Society.

The lowest fundamental modes of the two forms of transverse vibrations of a ring, such as the vibrations that were seen in the experiments, must be much graver than the elliptic vibration of the section. It is probable that the vibrations which constitute the incandescence of sodium-vapour are analogous to those which the smoke-rings had exhibited".

As examples of other type of problems are oscillations of sound and electromagnetic waves into various types of the closed cavities, whose surface is motionless. Such cavities refer to as closed wave-guides or resonators and consequently we will conditionally name this type of problems the closed wave-guide or resonator problems. In the classical physics a set of researches is devoted to such type of problems. Examples of such type of problems are also eigenvalues problems of wave functions in the quantum mechanics, which we will consider briefly below.

The above first and second type of problems leads to solutions of type of the standing waves, which have the relative time stability.

Thus, it is possible to assume, that stability (or the relative stability named metastability) of electromagnetic particles is connected with a formation of standing waves.

*As is known, a mathematical condition of occurrence of standing waves is the proportionality of wavelength to the size of body (volume), in which the wave propagates.* Therefore, at the search of a possible solution of these sorts of problems the basic role the limits play, which are imposed on propagation of waves or, in other words, the boundary states, imposed on wave functions.

Below we will show that from this boundary states follow the quantization conditions of characteristics of electromagnetic elementary particles.

## 5.1. Photon Wave Equation of CED

From Maxwell-Lorentz equations it is easy to obtain (Matveev, 1989) wave equation for the electric and magnetic field vectors:

$$\left(\frac{\partial^2}{\partial t^2} - c^2 \vec{\nabla}^2\right) F(\vec{r},t) = 0, \qquad (5.1)$$

where $\vec{F}$ is whichever of the EM wave functions.

$$\frac{\omega^2}{c^2} = k_x^2 + k_y^2 + k_z^2$$

The general harmonic solution of this wave equation has the complex $F(\vec{r},t) = F(\vec{r})e^{-i\omega t} = F_0 e^{i(\vec{k}\vec{r} - \omega t)}$ or trigonometric forms $F(\vec{r},t) = F_0 \cos(\vec{k}\vec{r} - \omega t)$, where $\omega = 2\pi\nu$ is the angular frequency, $\vec{k} = \frac{2\pi}{\lambda}\frac{\vec{p}}{|\vec{p}|}$ is the wave vector (here $\nu$ is the frequency, $k = |\vec{k}|$ called the *wave number*). Using these solutions it is also easy to obtain the dispersion law for EM waves:

$$\frac{\omega^2}{c^2} = k_x^2 + k_y^2 + k_z^2$$

We will see that the exponential form, although a complex number, proves more convenient.

Putting this solution in (5.1) we find for $F(\vec{r})$ the following equation for stationary waves:
$$(\vec{\nabla}^2 + k^2)F(\vec{r}) = 0, \quad (5.2)$$
where $k = \omega/\upsilon = 2\pi/\upsilon T = 2\pi/\lambda$, $T$ is the period and $\lambda$ - wavelength.

The equation (5.3) refers to as Helmholtz equation and is universal for the description of coordinate dependence of characteristics of harmonic waves.

Within the framework of this equation, is constructed the Kirchgoff diffraction and interference theory of light, which has excellently proved to be true an enormous experimental material.

**5.2. Wave equation solution for resonator**

To analyse the electromagnetic wave equation solution for resonator we will take (Wainstein, 1957) an orthogonal box from metal with *a*, *b* and *d* sites as our model of resonator. We will show that this solution is the standing electromagnetic waves.

According to (5.2) the electric field must satisfy the equations $(\vec{\nabla}^2 + k^2)\vec{E}(\vec{r}) = 0$ and $\vec{\nabla}\vec{E} = 0$ with the boundary state $\vec{E}_{II} = 0$ at the walls of the cavity (because inside the walls the electric energy will be rapidly dissipated by currents or polarization, the electric field intensity drops rapidly to zero into the walls). However, there could be an electric field *perpendicular* to the walls, because there could be surface charge on the wall. This means a possible solution is:

$$\vec{E}_x = E_{0x}k_x \cos k_x x \, \sin k_y y \, \sin k_z z$$
$$\vec{E}_y = E_{0y}k_y \sin k_x x \, \cos k_y y \, \sin k_z z, \quad (5.3)$$
$$\vec{E}_z = E_{0z}k_z \sin k_x x \, \sin k_y y \, \cos k_z z$$

For example, taking any $x$ for which $\sin k_x x = 0$, the second and third terms above are identically zero, but the first term certainly isn't.

Also from $div\vec{E} = 0$ using (5.3) we find $\vec{\nabla}\vec{E} = (E_{0x}k_x + E_{0y}k_y + E_{0z}k_z)\sin k_x x \, \sin k_y y \, \sin k_z z = 0$ if choosing $\vec{k}$ so that $\vec{k} \cdot \vec{E}_0 = 0$.

Here the wave equation requires $k_x = m\pi/a$, $k_y = n\pi/b$, $k_z = l\pi/d$, $\omega^2 = c^2(k_x^2 + k_y^2 + k_z^2)$ or $\omega = c\sqrt{k_x^2 + k_y^2 + k_z^2}$, where $(l,m,n)$ are positive integers, e.g. (1, 1, 0) or (3, 2, 4). In other words, each possible standing electromagnetic wave in the box corresponds to a point in the $(k_x, k_y, k_z)$ space labelled by three positive integers.

If we want also to obtain the general solution of the magnetic field, we first observe that the magnetic field satisfies the same equations and the boundary states as the electric field, and so the solution looks exactly the same as the electric solution. An alternative way is to use $\vec{B} = \vec{\nabla} \times \vec{E}/i\omega$, which can be easily obtained from Maxwell theory.

Thus, the character of the general solution for EM wave in the cavity is the standing electromagnetic wave.

It is easy to see, that the stated above description of occurrence of a resonance of the linear waves, if we make it in the complex form, will correspond to the occurrence of the resonance of the curvilinear (twirled) waves.

Show now that the quantum wave equation solutions for the stationary states give the identical results.

## 6.0. The quantum wave equations and their solutions for stationary waves

De Broglie has assumed that material particles alongside with corpuscular properties have as well the wave properties so that to the energy and momentum of a particle in a corpuscular picture there correspond the wave frequency and wavelength in a wave picture. De Broglie has shown that in this case from relativistic transformations the parities strictly follow:

$$\varepsilon = \hbar\omega \text{ and } \vec{p} = \frac{\hbar}{2\pi\lambda}\frac{\vec{p}}{|\vec{p}|} = \hbar\vec{k}$$

and the wave properties of a material particle are described by the same formula of a flat wave, as well as for an electromagnetic wave:

$$\psi(\vec{r},t) = \psi(\vec{r})e^{-i\omega t} = \psi_0 e^{i(\vec{k}\vec{r}-\omega t)} = \psi_0 e^{\frac{i}{\hbar}(\vec{p}\vec{r}-\varepsilon t)}$$

Thus the dispersion law for de Broglie wave it is easy to find from the energy-momentum conservation law for a particle:

$$\frac{\varepsilon^2}{c^2} = m_0^2 c^2 + \vec{p}^2$$

Really, replacing the energy and momentum by the wave characteristics, we will receive a dispersion correlation for waves of a matter:

$$\frac{\omega^2}{c^2} = \frac{m_0 c^2}{\hbar^2} + \vec{k}^2$$

It is easy to see, that within the framework of CWED this dispersion correlation satisfies to the equation of the twirled photon (see ), which produce the Dirac equations for the electron and positron.

We will consider now, to what wave equation there corresponds this dispersion correlation.

### 6.1. Helmholtz Equation for de Broglie waves

The Helmholtz equation (5.2) describes the waves of various nature in homogeneous mediums with constant frequency ($\omega = const$) and vacuum. The constancy of wavelength is not supposed.

Planck's correlation $\varepsilon = \hbar\omega$ shows that the condition $\omega = const$ entails the equality $\varepsilon = const$. Hence, Helmholtz equation can be applied to de Broglie waves at the description of motion of corpuscles in potential fields when their full energy is constant:

$$\varepsilon = \varepsilon_k + \varepsilon_p = p^2/2m + \varepsilon_p = const, \qquad (6.1)$$

where $\varepsilon_k = p^2/2m$ is a kinetic energy, $\varepsilon_p(\vec{r}) \equiv V(\vec{r})$ is potential energy of a corpuscle in a field. From a de Broglie correlation $\vec{p} = \hbar\vec{k}$ in view of (6.1) the equality follows:

$$k^2 = \frac{2m}{\hbar^2}(\varepsilon - \varepsilon_p), \qquad (6.2)$$

Substituting the expression (6.2) for $k^2$ in (5.3) we receive the equation:

$$\left(\vec{\nabla}^2 + \frac{2m}{\hbar^2}(\varepsilon - \varepsilon_p)\right)F(\vec{r}) = 0, \qquad (6.3),$$

named the Schroedinger stationary equation.

From this follows, that the existing calculation methods of the energy, momentum, angular momentum and other characteristics of particles state in the quantum field theory are calculations of resonance states of elementary particles in the various types of resonators, which in the quantum theory are usually named the potential wells. From the mathematical point of view these problems refer to as eigenvalues problems.

Consider the connection of these problems with CWED.

## 7.0. Principles of formation of elementary particles spectra in QED

The first calculations of quantum systems concerned the electron motion in the orbits of the hydrogen atom. The formulas of quantization of electron characteristics in the hydrogen atom have been first found empirically (formulas of Balmer, Paschen, etc.) . Then, it has been shown that they turn out as consequence of conditions of Bohr quantization.

The generalization of Bohr quantization rules has been made independently by Wilson and Sommerfeld. They have shown, that in case of systems with any number of degree of freedom it is possible to find such generalized coordinates $q_1, q_2, ..., q_f$, in which the motion of system is separated on $f$ harmonic oscillations; in this case a known rule of oscillator quantization can be applied for any of degrees of freedom. Owing to this generalization we receive $f$ quantum conditions:

$$\oint p_1 dq_1 = \left(n_1 + \frac{1}{2}\right)h, \quad \oint p_2 dq_2 = \left(n_2 + \frac{1}{2}\right)h, \quad ..., \quad \oint p_f dq_f = \left(n_f + \frac{1}{2}\right)h, \qquad (7.1)$$

where the integers $n_1, n_2, ..., n_f$ refer to as quantum numbers.

As an example of the application of these rules we will present the results the hydrogen-like atom calculation. Electron position in space at its motion around a nucleus is characterized by three polar coordinates $r, \vartheta, \psi$, describing radial, "equatorial" and "latitude" (azimuthal) motions accordingly. Therefore quantum states in this case become

$$\oint p_r dr = \left(n_r + \frac{1}{2}\right)h, \quad \oint p_\vartheta d\vartheta = \left(n_\vartheta + \frac{1}{2}\right)h, \quad \oint p_\psi d\psi = \left(n_\psi + \frac{1}{2}\right)h, \qquad (7.2)$$

The generalized momentums $p_r, p_\vartheta, p_\psi$ are calculated by the general rule: it is necessary at first to write the expression of kinetic energy in polar coordinates $r, \vartheta, \psi$:

$$\varepsilon_k = \frac{m}{2} v^2 = \frac{m}{2}\left(\dot{r}^2 + r^2 \dot{\vartheta}^2 + r^2 \sin^2 \vartheta \cdot \dot{\psi}^2\right), \qquad (7.3)$$

and to find the derivatives regarding the generalized velocities (which the corresponding momentums are):

$$p_r = \frac{\partial \varepsilon_k}{\partial \dot{r}} = m\dot{r}, \quad p_\vartheta = \frac{\partial \varepsilon_k}{\partial \dot{\vartheta}} = mr^2 \dot{\vartheta}, \quad p_\psi = \frac{\partial \varepsilon_k}{\partial \dot{\psi}} = mr^2 \sin^2 \vartheta \dot{\psi}, \qquad (7.4)$$

Then, using (7.2), it is possible to receive the quantization formulas of the momentums, defined by radial, "equatorial" and "altitude" quantum numbers: $n_r, n_\vartheta, n_\psi$

As de Broglie has shown, the Bohr or Wilson-Sommerfeld rules of quantisation define the conditions of the electron wavelengths integrality on various closed trajectories. Obviously, since any field can be represented as the oscillators sum, it is necessary to consider this rule as true for any quantum systems.

It is not difficult to see, that within the framework of CWED these rules are natural rules of a resonance of the twirled electromagnetic waves, if we take into account a quantization rule of their energy according to Planck-de Broglie.

The results, received according to Wilson-Sommerfeld quantization rules, have later appeared as solutions of the wave equation for standing de Broglie waves (i.e. the Schroedinger equation) for various sorts of potential wells (Shpolskii, 1951). Thus, Schroedinger equation is the equation for calculation of resonance states of an electronic wave in potential wells (resonators) of various type, boundary of wave motion in which are defined by potential energy of the system. Note, that the boundary states are expressed here by the same way, as in the classical theory of EM field:

$$\psi(a) = 0, \ \psi(b) = 0, \ \psi(d) = 0, \qquad (7.5)$$

It is easy to show, that this problem is absolutely identical to the problem of stand EM wave in resonators (and also to the problem of oscillation of strings, membranes or elastic body). The distinction is that the *wave vector is not constant here, but by some complex way depends on spatial coordinates; or, in other words, the dispersion relation is here defined by potential of system, which varies from a point to point according* (6.2).

Actually, it is not difficult to imagine that medium in the electromagnetic resonator can possess a dispersion, depending on spatial coordinates under the same law, as potential energy in a potential well of quantum-mechanical problem. Recollecting that within the framework of CWED the EM wave function is identical to wave function of quantum mechanics, it is easy to see, that boundary states in a quantum-mechanical problem must coincide with the boundary states in CED and CWED.

## Conclusion

Thus, we have shown, that the own spectra of elementary particles in CWED must arise in the same manner as the resonance states in any wave theory. The originality in comparison with calculation of stationary states of a particle in a field of other particles (solution for Schroedinger or Dirac electron equations) consists in the fact that in this case we have not an external field (i.e. an external potential box), but the particles themselves are like such a box.